Periodic Cluster Attractors and their Stabilities

in the Turbulent Globally Coupled Map Lattice

Tokuzo Shimada and Kengo Kikuchi

*Department of Physics, Meiji University, Higashi-Mita 1-1-1, Kawasaki, Kanagawa 214-8571, Japan*

**Abstract**

The Globally Coupled Map Lattice (GCML) is one of the basic model of the intelligence activity. We report that, in its so-called turbulent regime, periodic windows of the element maps foliate and systematically control the dynamics of the model. We have found various cluster attractors. In one type of them, the maps split into several almost equally populated clusters and the clusters mutually oscillate with a period (*p*) that is the same with the number of clusters (*c*). We name them as maximally symmetric cluster attractors (MSCA's). The most outstanding are the *p3c3* MSCA and its bifurcate. The MSCA is proved to be linearly stable by Lyapunov analysis. There are also cluster attractors with *p*>*c*. They come out in sequences with increasing coupling. The formation of the clustors in the very weakly coupled chaotic system may suggest a new form of an intelligence activity.

**1. Introduction**

We have recently found that, in the so-called turbulent regime of the GCML, a variety of amazing periodic cluster attractors are formed even though the coupling between the element maps is set to be very small [1]. In this note we review our work and substantiate it by ample examples. For a pioneering work of this model, see e.g. Ref. [2].

The simplest GCML—a homogeneous one—is defined by

$$x_i(t+1) = (1-\varepsilon) f_a(x_i(t)) + \varepsilon h(t), \tag{1}$$

$$h(t) = \sum f_a(x_j(t))/N, \tag{2}$$

with $f_a(x) \equiv 1 - ax^2$. All maps are endowed with a common high nonlinearity and evolve under an averaging interaction via their mean field $h(t)$ with a coupling $\varepsilon$. The formation of clusters via synchronization in this model is well studied for the large coupling region and the switch between coded attractors has been investigated in detail. For a recent progress, see [3]. On the other hand, in the small coupling (so called turbulent) region, maps have been regarded to evolve almost randomly under some hidden-coherence. Our new observations of various cluster formation in this region indicate that a pattern-recognition in a complex system is possible even if a highly random system is set with a very weak coupling.

We establish the linear stability of the cluster attractors formed in the turbulent regime. We in particular derive algebraically the $\varepsilon$ value for the formation of the MSCA.

**2. Foliation of the Element Map Windows**

In a MSCA configuration the mean field must be time-independent due to a high symmetry in the cluster populations. Thus (1) becomes

$$x_i(t+1) = (1-\varepsilon) f_a(x_i(t)) + \varepsilon h^*, \tag{3}$$

with a constant external field $h^*$. The evolution equation acts on all the maps commonly in (1) and it becomes furthermore independent of time in (2). We can cast this unique equation into a standard logistic map with a reduced nonlinear parameter $b$,

$$y_i(t+1) = 1 - b(y_i(t))^2, \tag{4}$$

by a linear scale transformation

$$y_i(t) = (1-\varepsilon+\varepsilon h^*)^{-1} x_i(t), \tag{5}$$

and the reduction factor of nonlinearity is

$$r \equiv b/a = (1-\varepsilon)(1-\varepsilon(1-h^*)). \tag{6}$$

The clusters of MSCA oscillate mutually around the fixed average $h^*$. Their orbits are the same each other *modulo time translation* and propor-tional to the orbit of a logistic map with the nonlinearity $b$ reduced by a factor in (6). Thus it must hold that

$$y^* = (1 - \varepsilon + \varepsilon h^*)^{-1} h^*, \tag{7}$$

with $y^*$ being the time-average of the logistic map at $b$. Equation (6) and (7) are the key to find how a periodic window of an element map foliates and produces a MSCA of GCML. On one hand there is a periodic window at $b$ with $y^*(b)$. On the other hand there is a MSCA produced in GCML at $a, \varepsilon$ with a constant mean field $h^*$. The nonlinearity $a$ of the latter element map is reduced to $b$ by the averaging interaction. For each reduction factor $r$, we can work out $a$ and $\varepsilon$ that corresponds to $b$ by eliminating $h^*$,

$$\begin{aligned} a^{(b)}(r) &= b/r, \\ \varepsilon^{(b)}(r) &= 1 - ry^*(b)/2 - \sqrt{r(1-y^*(b)) + (ry^*(b)/2)^2}. \end{aligned} \tag{8}$$

Now by varying $r$, (8) gives *a curve of balance* on the $a,\varepsilon$-plane, which emanates from the point $(b,0)$. We call this curve as ***a foliation curve*** of a window dynamics. *If a MSCA with a period p is to be produced, it must be produce in a GCML with the parameter a, ε set on the curve of the period p window. At some stronger coupling at given a, the maps should be more tightly bunched and we may expect $p > c$ type attractors.*

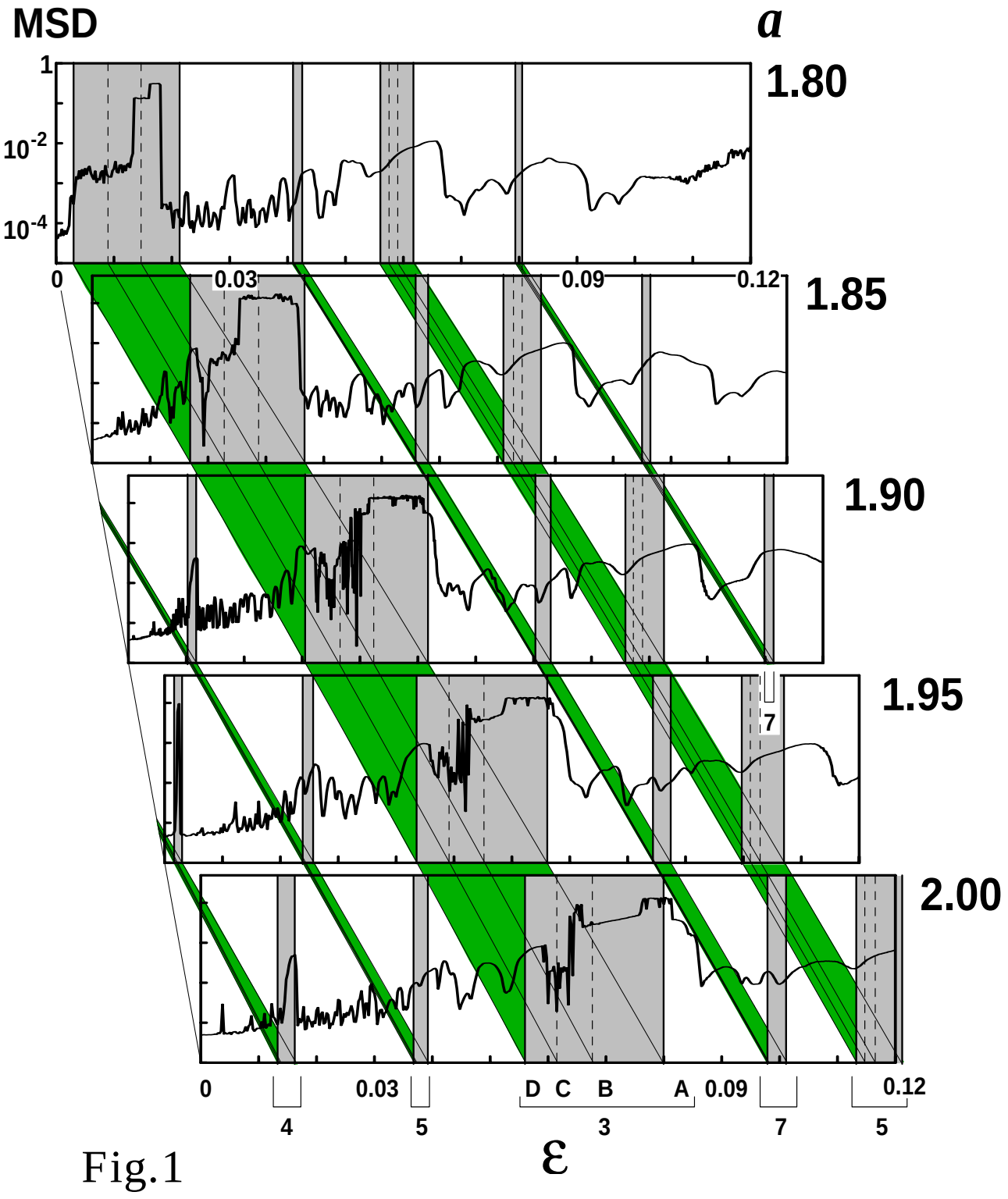

Fig.1

Let us check if this prediction works. For this purpose we consider the mean squared deviation (MSD) of the mean field in time

$$\langle \delta h^2 \rangle_T \equiv \Sigma_t (h(t) - \langle h \rangle_T)^2 / T, \tag{9}$$

as an indicator of the cluster formation. A MSCA will yield very low MSD due to its high symmetry in cluster populations, while the $p > c$ attractors will give remarkable peaks due to the lack of one or more clusters. In Fig.1 each panel is set at a fixed $a$ and MSD is shown as a function of $\varepsilon$ for $N$=$10^4$ GCML. Six prominent logistic widows ($p = 7, 5, 7, 3, 5, 4$ with increasing $b$ —decreasing $\varepsilon$ at the same $a$) are selected and the family of foliation curves of these windows is shown underneath the panels. The four curves A-D for each window respectively come from the point A below the threshold, the threshold B, the first bifurcation point in the window C, and the closing point D. The shaded zones in each panel are then the expected place of the manifestation of prominent windows. At each zone, a MSD valley due to MSCA should appear in the lower $\varepsilon$ side and a MSD peak by $p > c$ cluster attractors at the nearby higher $\varepsilon$. We find that the prediction works with almost no failure in all panels and in all six windows.

Interestingly, the MSD curve in each panel has an ample amount of peaks and valleys at the smaller $\varepsilon$ region (the left), but only a few broad ones at the larger $\varepsilon$. This is naturally understood as follows. In a way, *each panel is a screen which displays the windows of the single logistic map by using a macroscopic coherent state of GCML.* But the panels are inclined; a smaller $\varepsilon$ implies less reduction, i.e. $r \approx 1$. Hence the left sensitively displays the sharp peak-valley structure induced by cluster attractors. The right, on the other hand, can reflect only the accumulation of the periodicity remnants from nearby windows, being dominated by the prominent one at its respective zone.

We have checked that, at all the MSD valleys with large nonlinearity reduction, the $h(t)$ distribution is Gaussian with the MSD sizably larger than the value dictated by the law of large numbers—the so-called hidden coherence [2]. *We therefore conjecture that the hidden coherence is due to a desynchronized MSCA state* [1].

## 3. Lyapunov Stability Analysis

The Lyapunov analysis can be applied to both diverging and converging system orbits and it can detect the possible coexistence of multifold finial states. It tracks the expansion rate of a shift vector under the linearized GCML equation

$$\delta x_i(t+1) = -2a[(1-\varepsilon+\frac{\varepsilon}{N})x_i(t)\delta x_i(t) + \frac{\varepsilon}{N}\sum_{i\neq j} x_j(t)\delta x_j(t)], \qquad (10)$$

and yield the maximum exponent $\lambda_{\max}$.

First let us investigate the outstanding period three cluster attractors. In Fig.2 we compare $\lambda_{\max}$ and the MSD. $N$=10[6], $a$ =1.90.

We find three salient structures.

**(1) MSCA*.** *A seagull structure (ε=0.032-0.037)* in both. *For the most stable events, the MSD is also the least.* By a direct observation of the orbits, we find all events are bifurcated MSCA. This can be understood as follows. The maps and the mean field together are a bootstrap system, see (1). That is, generally the mean field is not a simple external source and the fluctuation in it will be reflected to the fluctuation of the maps. The high mean field fluctuation would lead to the instability of the system. An exception is the MSCA. Here the mean field is constant and the system protects itself from instability which is otherwise amplified by the bootstrap. *The MSCA is the configuration with which the GCML stabilizes itself with minimum fluctuation of the mean field.*

**(2) p3c3 MSCA.** *The first low band (0.037-0.041).* The negative $\lambda_{\max}$ and the low MSD.

**(3) p3c2 cluster attractor.** *The second low band (0.041-0.051).* Here, the MSD is extremely high because of a lack of one cluster to minimize the fluctuation. For the bulk of events we find $\lambda_{\max}$ is small but positive. For a system with low degrees of freedom, the positive $\lambda_{\max}$ implies chaos. But here, even with a positive $\lambda_{\max}$, the maps always form stable *p3c2* state. There is actually no contradiction. The global motion of the clusters is periodic, but, inside each cluster, maps are evolving randomly. The Lyapunov exponent is sensitive to the microscopic motion and hence yields positive value. But for a larger deviation, nonlinear terms can become relevant and pull back the map. *This type of map motion—microscopically chaotic but macroscopically in the periodic clusters—may be called as confined chaos.*

Note that in Fig.2 there are also states with high rate mixing of maps (denoted as M), which coexist with the cluster attractors at the same *ε*.

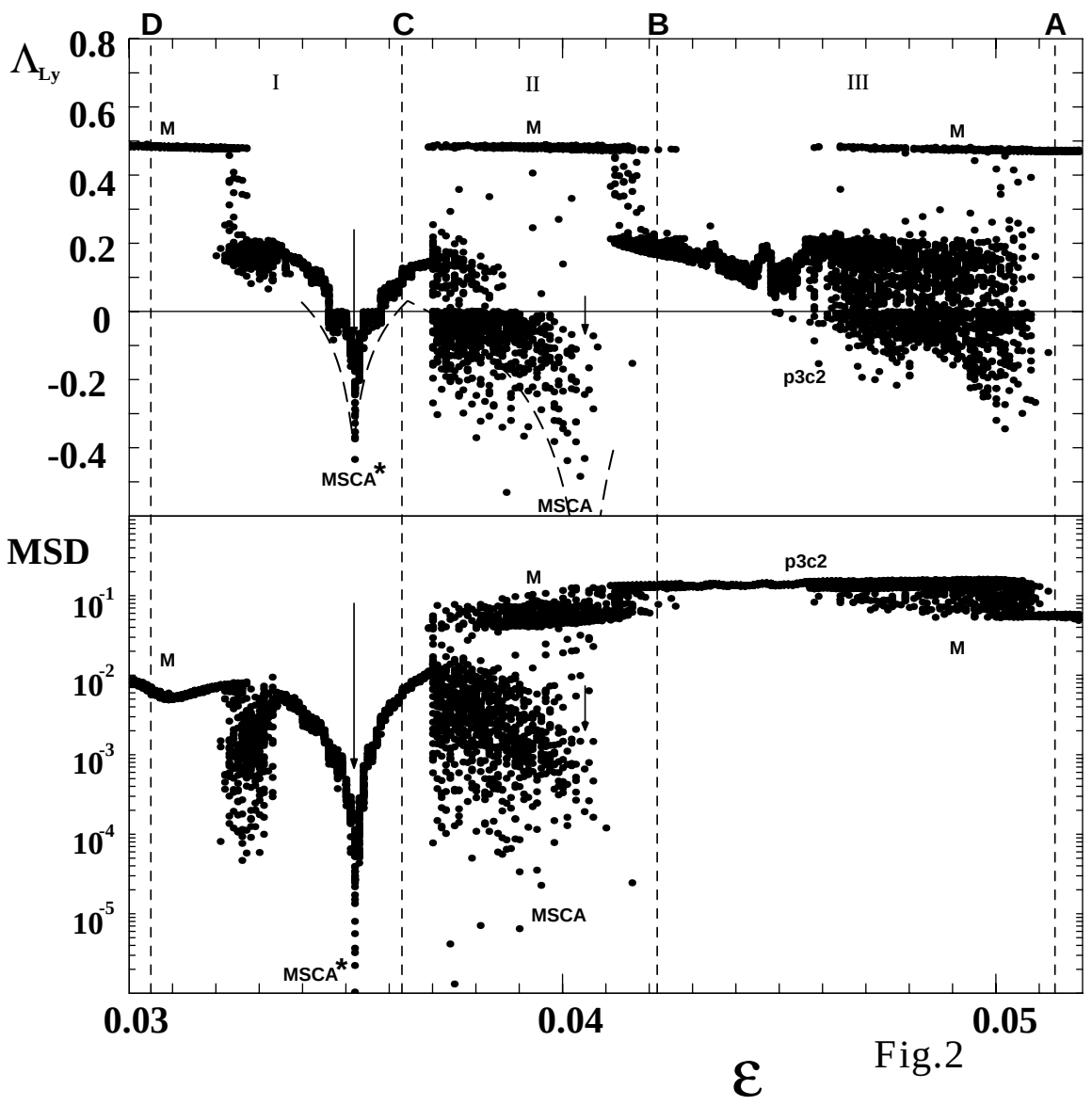

Fig.2

Let us predict the position of the salient cusp. For a cluster attractor configuration, there occurs a high degeneracy of eigenvalues of the linear stability matrix. For each cluster ($I$) with $N_I$ maps, there is a single eigenvalue with ($N_I$-1)-fold degeneracy. It is

$$\Lambda^{(I)} = (-2a(1-\varepsilon))^p \Pi_{k=1}^{p} X_k^I, \qquad (11)$$

where $X^I$ denotes the cluster orbit. It is responsible for the stability of a map in that cluster. The altogether $c$ eigenvalues of this type take care of $\Sigma_{I=1}^{c}(N_I-1) = N-c$ degrees of freedom of maps. The other $c$ eigenvalues are responsible for the stability of the cluster orbits. For a MSCA the product of $X_k^I$ over the period $p$ is common to $I$. Therefore the spectrum consists of a highly degenerate eigenvalue $\Lambda$ with $(N-c)$-fold degeneracy and additional $c$ non-degenerate eigenvalues.

Now, the crucial point. The orbit $X_k^I$ of a MSCA cluster at *a, ε* (common to all $I$) is nothing but the orbit of a single map $y_k$ at $b$ modulo a scale factor (see (5)). *Thus,* $\Lambda$ *for MSCA agrees with the Lyapunov exponent of a single map at* $b$ *;*

$$\Lambda = (-2b)^p \Pi_{k=1}^{p} y_k. \qquad (12)$$

*This becomes zero when one of the orbit points* $y_k$*'s becomes zero, that is, when* $b$ *is a solution of*

$(f_b)^p(0)=0$. As for the other $c$ eigenvalues, they agree with $\Lambda$ within small correction of order $\varepsilon/c$ [1]. That is, if $\lambda_{\max}=\log(|\Lambda|)/p$ is $-\infty$, they are all extremely small, approximately $\log(\varepsilon/c)$. For *p6c6* cusp, the relevant solution $b_{c6}$ is 1.77289. The foliation curve for this $b$ reaches the $a=1.90$ panel at $\varepsilon_{c6}$=0.0352. This is precisely the MSCA* cusp position. The predicted curve for $\lambda_{\max}$ around the cusp (the dashed line) is also in good agreement with the data.

Now let us investigate the cluster attractors with higher periodicities.

***p5* and *p4* cluster attractors**

In Fig.3a we show the same with Fig.2 for the foliation of the *p*5 window—the sequence of $MSCA^* \to c5 \to c4 \to c3 \to c2$ sampled at $a=1.64$. Fig.3b is for *p4;* $MSCA^* \to c4 \to c3 \to c2$ at $a=1.95$. Both for $N=10^4$. Overall agreement with the *p*3 case can be seen clearly—we observe the MSCA cusps in D-B and $p>c$ clusters in B-A. Algebraically we obtain $b_{c10}$=1.62943 and $b_{c8}$=1.94178. The predicted MSCA* cusp positions from our foliation equations are $\varepsilon_{c10}$=0.00397 and $\varepsilon_{c8}$=0.00194—in agreement with the observed ones.

In Fig.4 we show the composition of $p>c$ attractors obtained by a gap analysis of $10^3$ random events at each $\varepsilon$ for each of *p*5 and *p*4. With increasing $\varepsilon$, the $c$ sequentially decreases with intermediate coexisting phases of $p>c$ clusters and random maps. The MSCA dynamics dominates inside the window (above the dashed line B) while the $p>c$ clusters are formed in the intermittent region below the window (the region of the higher reduction by $\varepsilon$).

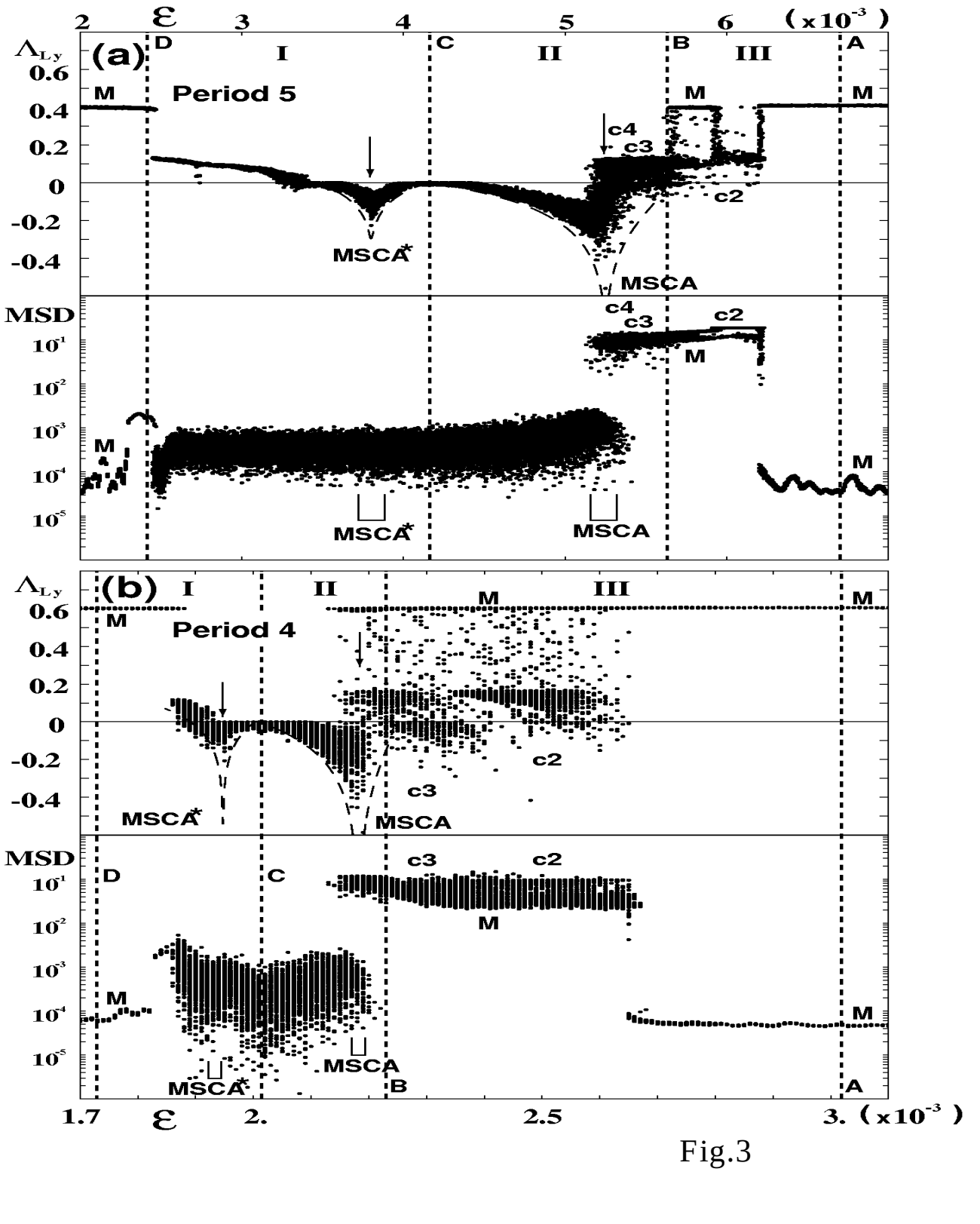

Fig.3

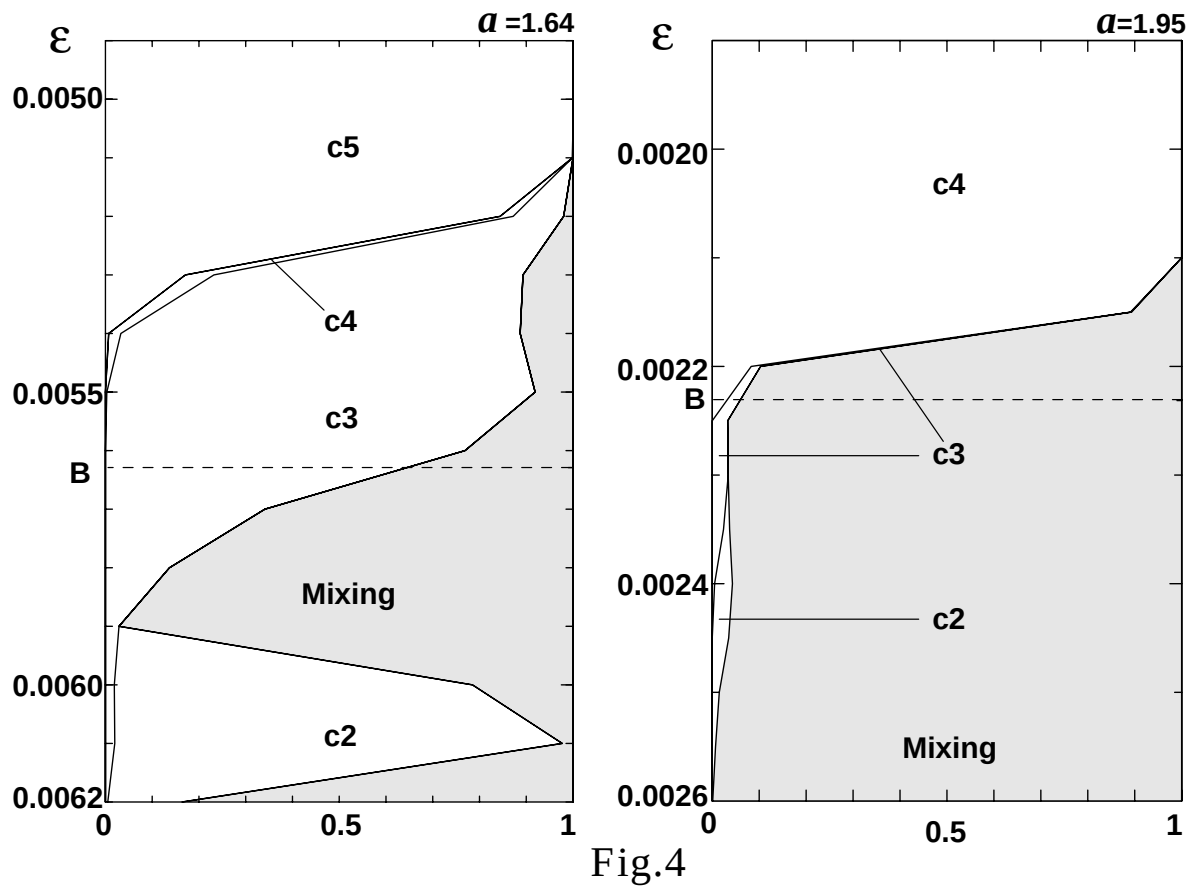

Fig.4

## 4. Conclusion

The newly found stable self-organized MSCA's (and its bifurcates) with a minimum fluctuation are the basic states of maps in the turbulent GCML. They may be regarded as counter parts of the ordered vacuum in the field theory at the spontaneously broken symmetry phase. The $p>c$ attractors are curious deformed states at slightly higher coupling. Their periodic orbits are almost the same with the MSCA but due to the lack of some clusters the MSD is maximized. Even when the Lyapunov exponent is positive, the maps are macroscopically confined in clusters stably.

The foliation has been also found independently by two other groups [4] but neither the stability nor the $p>c$ attractors were discussed by them.

---

[1] T. Shimada, Tech. Rep. IEICE, NPL97-159, 71(1998); T. Shimada and K. Kikuchi, Phys. Rev. **E**62, 3489 (2000).

[2] K. Kaneko, Phys.Rev.Lett. **63**, 219 (1989); ibid. **65**, 1391 (1990).

[3] T. Shimada and S. Tsukada, *A resolution of the puzzle of the posi-nega switch mechanism in the globally coupled map lattice*, submitted to this conference (AROB01-6).

[4] T. Shibata and K.Kaneko, Physica D **124**, 177 (1998). A.P. Parravano and M. G. Cosenza, Int. J.Bifurcation Chaos **9**, 2331(1999).